\documentclass[11pt]{elsart}
\usepackage{amsfonts}
\usepackage{slashbox}
\usepackage{graphicx}
\usepackage{amssymb,amsmath}
\usepackage{mathrsfs} 

\begin{document}
\begin{frontmatter}
\title{A note on hierarchies and bureaucracies}
 \author{Shang-Guan H. Wu\corauthref{cor}}
\corauth[cor]{Corresponding author.} \ead{hywch@mail.xjtu.edu.cn}
\address{Wan-Dou-Miao Research Lab, Suite 1002, 790 WuYi Road,\\
Shanghai, 200051, China.}
\begin{abstract}
In this note, we argue that there is a bug in [Tirole, J.,
``Hierarchies and bureaucracies: On the role of collusion in
organizations,'' {\em Journal of Law, Economics and Organization},
vol.2, 181-214, 1986].
\end{abstract}

\begin{keyword}
Collusion; Incentive theory.
\end{keyword}
\end{frontmatter}

In this note, the notation is referred to Ref. [1]. Vertical
structures are represented by three-layer hierarchies:
principal/supervisor/agent. The \emph{principal}, who is the owner
of the vertical structure or the buyer of the good produced by the
agent, or more generally, the person who is affected by the agent's
activity, lacks either the time or the knowledge required to
supervise the agent. The \emph{supervisor} is a party that exerts no
effort, receives a wage from the principal and collects information
to help the principal control the agent. The \emph{agent} is the
productive unit. The profit $x$ created by the agent's activity
depends on a productivity parameter $\theta$ and on the effort $e>0$
he exerts:
\begin{equation*}
  x=\theta+e.
\end{equation*}
The agent's effort $e$ is assumed not observable by the supervisor
and the principal. The agent's disutility of effort is equal, in
monetary terms, to $g(e)$.

For a given $\theta$, the supervisor's signal $s$ can take two
values: $\{\theta,\varnothing\}$, where $\varnothing$ denotes
observation ``nothing''. The report $r$ of supervisor is
\emph{verifiable}, i.e., if $s=\theta$ then $r\in
\{\theta,\varnothing\}$, otherwise $r=\varnothing$. The productivity
$\theta$ can take two values: $\underline{\theta}$ and
$\bar{\theta}$ such that $0<\underline{\theta}<\bar{\theta}$. There
are four states of \emph{nature}, indexed by $i$. State of nature
$i$ has probability $p_{i}$ ($\sum\limits_{i=1}^{4}p_{i}=1$). The
agent always observes $\theta$ before choosing his effort. The
supervisor may or may not observe $\theta$. In the following
description of the four states of nature, $S$ and $A$ stand for
supervisor and agent:\\
\emph{State 1}: $A$ and $S$ observe $\underline{\theta}$.\\
\emph{State 2}: $A$ observes $\underline{\theta}$, $S$ observes
``nothing''.\\
\emph{State 3}: $A$ observes $\bar{\theta}$, $S$ observes
``nothing''.\\
\emph{State 4}: $A$ and $S$ observe $\bar{\theta}$.

\textbf{Claim 1}: The supervisor and principal cannot discriminate
\emph{State} 2 and 3.\\
\emph{Proof}: The only difference between \emph{State} 2 and 3 is
that agent $A$ observes different values of productivity. However,
this parameter is agent's private information and not observable to
the supervisor and principal. Put in other words, \emph{State} 2 and
3 are indifferent to the supervisor and the principal. $\square$

\emph{Timing}.\\
1) The principal offers a contract.\\
2) $A$ learns the productivity $\theta$, $S$ learns the signal $s$.\\
3) $A$ chooses the effort $e$.\\
4) Profit $x=\theta+e$, $S$ reports $r$.\\
5) The principal transfers $S(x,r)$ and $W(x,r)$ to the supervisor
and agent respectively.

\textbf{Claim 2}: There is a bug in the agent's incentive
compatibility constraint (\emph{AIC}): $W_{3}-g(e_{3})\geq
W_{2}-g(e_{2}-\triangle\theta)$ (Page 191, Line 8, [1]).

\emph{Proof}: As specified in the timing, the wage $W(x,r)$ of agent
is transferred by the principal. It only depends on the commonly
observable variables $x$ and $r$, not on the agent's private
productivity $\theta$. Since the principal cannot discriminate
\emph{State} 2 and 3, the items $W_{2}$ and $W_{3}$ are indeed
meaningless.

That's the bug, not only for the condition (\emph{AIC}), but also
for the whole paper of Tirole (1986).

\section*{Acknowledgments}

The author is very grateful to Ms. Fang Chen, Hanyue Wu
(\emph{Apple}), Hanxing Wu (\emph{Lily}) and Hanchen Wu
(\emph{Cindy}) for their great support.


\end{document}